\RequirePackage{fix-cm}
\documentclass{svjour3}                     % onecolumn (standard format)
\smartqed  % flush right qed marks, e.g. at end of proof
\usepackage{graphicx}
\usepackage{bm}
\usepackage{here}
\usepackage{tikz}
\begin{document}

\title{Calculations of magnetic field produced by spin-vortex-induced loop currents in Bi$_2$Sr$_2$CaCu$_2$O$_{8+\delta}$ thin films using the particle-number conserving Bogoliubov-de Gennes formalism %\thanks{Grants or other notes
}

%\titlerunning{Calculations by the particle number conserving BdG quations}        % if too long for running head

\author{ Hiroyasu Koizumi \and Haruki Nakayama \and Hayato Taya
       %etc.
}


\institute{H. Koizumi\at
              Center for computational sciences, University of Tsukuba, Tsukuba, Ibaraki, Japan \\
              Tel.:+81-29-8536403\\
              Fax:+81-29-8536403\\
              \email{koizumi.hiroyasu.fn@u.tsukuba.ac.jp}           %  \\
               \and
           H. Nakayama \at
              School of Science and Engineering, University of Tsukuba, Tsukuba, Ibaraki, Japan
               \and
           H. Taya \at
              Graduate School of Pure and Applied Sciences, University of Tsukuba, Tsukuba, Ibaraki, Japan
}

\date{Received: date / Accepted: date}

\maketitle

\begin{abstract}
A theory for cuprate superconductivity predicts the existence of nano-sized loop currents called, `` spin-vortex-induced loop currents (SVILCs)''. We calculate magnetic fields produced by them  for a model of Bi$_2$Sr$_2$CaCu$_2$O$_{8+\delta}$ (Bi-2212) thin films composed of one surface and two bulk CuO$_2$ bilayers.
In this model, bulk CuO$_2$ layers host stable spin-vortices around small polarons formed from doped holes; they give rise to a $U(1)$ gauge field described by the Berry connection from many-body wave functions, and generates the SVILCs.
The effect of the gauge field is taken into account by the particle-number conserving Bogoliubov-de Gennes (PNC-BdG) formalism. 
The magnitude of the calculated magnetic field produced by the SVILCs in the vicinity of the surface ($10a \approx 4$ nm, where $a$ is the lattice constant of the CuO$_2$ plane) is in the order of mT; thus, may be detectable by currently available detection methods. The detection of the SVILCs by the magnetic field measurement may bring about the elucidation of the cuprate superconductivity, and may also lead to their quantum device applications, including qubits.
\end{abstract}

\section{Introduction}

Since the discovery of high temperature superconductivity in 1986 \cite{Muller1986}, 
extensive efforts have been devoted to elucidate its mechanism. However, no widely-accepted theory exists despite more than 30 years of efforts.
Since the cuprate superconductivity is markedly different from the BCS superconductivity, the elucidation of it will require a marked departure from the standard superconductivity theory.

Besides, revisiting of experimental facts of superconductivity has found several problems in the standard theory: 1) the standard theory relies on  the use of particle number non-conserving formalism although superconductivity occurs in an isolated system where the particle-number is conserved \cite{Peierls1991}; 2) the superconducting carrier mass obtained by the London moment experiment is the free electron mass $m_e$, although the standard prediction is the effective mass $m^\ast$ of the normal state \cite{Hirsch2013b}; 3) the reversible superconducting-normal phase transition in a magnetic field cannot be explained by the standard theory \cite{Hirsch2017}; 4) the dissipative quantum phase transition in a Josephson junction system predicted by the standard theory is absent \cite{PhysRevX2021a}; 5) so-called the `quasiparticle poisoning problem' indicates the existence a large amount of excited single electrons in Josephson junction systems, obtaining the observed ratio of their number to the Cooper pair number  $10^{-9} \sim 10^{-5}$ in disagreement with the standard theory predicted value $10^{-52}$\cite{poisoning2023,Serniak2019}.
The existence of the above problems seems to indicate the need for serious revisions of the superconductivity theory.
It is sensible to consider that the theory explains the cuprate superconductivity will also need to resolve the above problems.

One of the present authors has put forward a new theory of superconductivity that encompasses the BCS theory and lifts the disagreements mentioned above  \cite{koizumi2022,koizumi2022b,koizumi2023}.
In this theory, supercurrent is generated by an emergent $U(1)$ gauge field 
arising from the singularities of the many-body wave function that can be detected by the Berry phase formalism \cite{Berry}. 
Especially, such a gauge field arises when spin-twisting itinerant motion of electrons is realized; 
in this case, singularities of the wave function exist at the centers of the spin-twisting, and the  emergent gauge field gives rise to persistent loop currents around them.

The cuprate superconductivity may be elucidated by the new theory mentioned above since the presence of coherence-length-sized spin-vortices and accompanying loop currents ({\em spin-vortex-induced loop currents} or {\em SVILCs}) are highly plausible from the following facts \cite{Koizumi2011,koizumi1,koizumi2,koizumi3,koizumi4,KoizumiDrude}: 1) the superconducting transition temperature for the optimally doped cuprates corresponds to the stabilization temperature of the coherence-length-sized loop currents \cite{Kivelson95}, and the experiment using Bi$_2$Sr$_2$CaCu$_2$O$_{8+\delta}$ (Bi-2212) thin films has confirmed that the superconducting transition is the BKT type \cite{D3RA02701E}; 2) theoretical calculations based on the stabilization of the SVILCs yields a reasonable transition temperature \cite{HKoizumi2015B,Koizumi2017}; 3)
the magnetic excitation spectra observed  may be taken as the evidence for the existence of nano-sized spin-vortices \cite{Neutron,Hidekata2011}; 4) the presence of SVILCs explains the polar Kerr effect measurement \cite{Kerr1}, enhanced Nernst effect measurement \cite{Nernst}, and the neutron scattering measurement \cite{neutron2015}. 
Recently, spin-vortices have been observed in the cupare superconductors \cite{Wang:2023aa}.
Although the observed spin-textures are different from the predicted spin-vortices, experiments by improved spatial resolution and sensitivity may detect the predicted nano-sized spin-vortices. 

In the present work, we calculate magnetic fields produced by the SVILCs for the purpose of 
detecting the SVILCs by measuring the magnetic field produced by them. 
The confirmation of the existence of the SVILCs will lead to the elucidation of the cuprate superconductivity.
Further, it also helps to achieve novel quantum device applications of cuprate superconductors.
Now, methods for preparing Bi-2212 thin films have been established \cite{WANG201213,Jiang:2014aa,Jindal:2017aa,adfm.201807379,SHEN202135067,nwac089,KEPPERT2023157822}.
Using those methods, qubits made of the SVILCs may be realized \cite{WAKAURA201655,Wakaura2017,Koizumi:2022aa}. 

The organization of the present work is as follows: In Section~\ref{Bog}, the particle number conserving Bogoliubov-de Gennes (PNC-BdG) formalism is briefly explained; this formalism is used to deal with the emergent gauge field generated by the spin-twisting itinerant motion of electrons \cite{koizumi2019,koizumi2021,koizumi2021b,Koizumi2021c}. 
 In Section~\ref{sec3}, the PNC-BdG equations for the surface plus bulk CuO$_2$ layers model
for Bi-2212 thin films are explained. In Section~\ref{sec4}, calculated results for SVILCs and
their magnetic fields are shown.
 Lastly, we conclude the present work in Section~\ref{sec5}. 

\section{Superconducting state described by the particle-number conserving Bogoliubov-de Gennes(PNC-BdG) formalism}
  \label{Bog}

 In the PNC-BdG formalism, the electron field operators for the superconducting state are given by
\begin{eqnarray}
\hat{\Psi}_{\uparrow}({\bf r})&=&\sum_{n} e^{-{i \over 2}\hat{\chi} ({\bf r})}\left( \gamma_{{n} } u_{n \uparrow}({\bf r})  -\gamma^{\dagger}_{{n} } v^{\ast}_{n \uparrow}({\bf r}) \right)
\nonumber
\\
\hat{\Psi}_{\downarrow}({\bf r})&=&
\sum_{n} e^{-{i \over 2}\hat{\chi} ({\bf r})} \left( \gamma_{{n} } u_{n \downarrow}({\bf r}) +\gamma^{\dagger}_{{n} } v^{\ast}_{n \downarrow}({\bf r}) \right)
\label{eq1}
\end{eqnarray}
where $u_{n \sigma}({\bf r})$ and $v_{n \sigma}({\bf r})$ are the single-particle basis functions with spin $\sigma$, $\uparrow$ and $\downarrow$ indicate up and down electron spin states, respectively, and $\gamma_{n }$ and $\gamma^\dagger_{n}$ are the particle-number conserving Bogoliubov operators that act in the same way as the usual Bogoliubov operators, but, conserve the particle-number.

The above electron field operators should be compared with the usual one,
\begin{eqnarray}
\hat{\Psi}_{\sigma}({\bf r})&=&\sum_{n} \varphi_{n \sigma}({\bf r})c_{n \sigma}
\end{eqnarray}
where $\{ \varphi_{n \sigma}({\bf r}) \}$ is the single-particle basis, and $c_{n \sigma}$ is the annihilation operator for the state described by $\varphi_{n \sigma}({\bf r})$.
In Eq.~(\ref{eq1}), the Bogoliubov operators $\gamma_{n }$ and $\gamma_{n }^\dagger$ are used instead of the electron annihilation operators, $c_{n \sigma}$, and the superconducting ground state, $|{\rm Gnd}(N) \rangle$, is defined by
\begin{eqnarray}
\gamma_{n }|{\rm Gnd}(N) \rangle=0
\label{eq2}
\end{eqnarray}
where $N$ denotes the  total particle-number of the system.
This indicates that the superconducting ground state is the vacuum of the Bogoliubov quasiparticle excitations. This formalism was first adopted by de Gennes \cite{deGennes} (see also Ref.~\cite{Zhu2016} for updated treatments), and the particle-number conserving one was put forward by one of the present authors in Ref.~\cite{Koizumi2021c}.

In the PNC-BdG, the operators $e^{\pm {i \over 2} \hat{\chi}({\bf r})}$ are also introduced.
They are number changing operators satisfying
\begin{eqnarray}
e^{\pm {i \over 2} \hat{\chi}({\bf r})}|{\rm Gnd}(N) \rangle= e^{\pm{i \over 2} {\chi}({\bf r})}|{\rm Gnd}(N\pm1) \rangle
\label{eqBC}
\end{eqnarray}
The change of the particle-number by minus one is achieved by the factor $e^{-{i \over 2} \hat{\chi}({\bf r})}$ in Eq.~(\ref{eq1}); and the Bogoliubov operators $\gamma_{n}$ and $\gamma_{n }^\dagger$ conserve the particle-number.
Note also that a phase factor $e^{\pm{i \over 2} {\chi}({\bf r})}$ appears in addition to the number change in the right-hand-side of Eq.~(\ref{eqBC}). This phase factor gives rise to the Berry phase
\begin{eqnarray}
\langle {\rm Gnd}(N)|e^{ {i \over 2} \hat{\chi}({\bf r}_f)}e^{-{i \over 2} \hat{\chi}({\bf r}_k)}|{\rm Gnd}(N) \rangle
=e^{ {i \over 2} \int_{{\bf r}_k}^{{\bf r}_f} \nabla {\chi} \cdot d{\bf r}}
\end{eqnarray}
arising from the emergent gauge field (or the Berry connection from many-body wave functions). In the present case, the gauge field arises from
the singularities of the many-electron wave functions existing at the centers of spin-twisting.
The number changing operators are obtained from the quantization of the collective mode described by $\chi({\bf r})$. Consult Refs.~\cite{koizumi2022,koizumi2022b} for the detail.

\section{Surface-bilayer plus Bulk-bilayers model}
  \label{sec3}

Let us explain the model for Bi-2212 thin films used in the present work. In this model, we only retain the sites for copper atoms in CuO$_2$ planes as in our previous work \cite{Koizumi2022aa}. Since the Bi-2212 consists of CuO$_2$ bilayers, we need to extend our previous model to incorporate this fact.

Our model Hamiltonian $H_{\rm eff}$ is given by
  \begin{eqnarray}
H_{\rm eff}
&=&H^{\rm HF}_{\rm sbilayer}+\sum_{\ell_b} H^{\rm HF}_{\rm bbilayer}(\ell_b)+H^{\rm HF}_{\rm surface-bulk}+\sum_{\ell_b}H^{\rm HF}_{\rm bulk-bulk}(\ell_b, \ell_b +1)
\nonumber
\\
&=&
\sum_{k,j,\sigma, \sigma'} h_{k \sigma, j \sigma'} c_{k \sigma}^{\dagger} c_{j \sigma'}
  + \sum_{k,j} [ \Delta_{kj} e^{ -{i \over 2} \hat{\chi}_j}e^{ -{i \over 2} \hat{\chi}_k}c_{k \uparrow}^{\dagger}c_{j \downarrow}^{\dagger}+ \mbox{H.c.}]
  +E_{\rm const}
  \nonumber
  \\
  \label{Heff}
  \end{eqnarray}
  where $k,j$ are site indices and $\sigma, \sigma'$ are spin indices; $E_{\rm const}$ is a constant arising from the mean field treatment; $c_{k \sigma}^\dagger$ and  $c_{k \sigma}$  denote
  the creation and annihilation operators for electrons with spin $\sigma$ at site $k$, respectively; $\ell_b$ is the index for the bulk bilayers. We consider the system with two bulk bilayers in the present work.

In the following, we just explain modifications made for the present model from the one used in Ref.~\cite{Koizumi2022aa}. The first term in $H_{\rm eff}$, $H^{\rm HF}_{\rm sbilayer}$, is the surface bilayer Hamiltonian consists of two CuO$_2$ layers connected by an interlayer hopping Hamiltonian. In the surface layers, the following pair potential $\Delta_{kj}$  given by
  \begin{eqnarray}
  \Delta_{kj}=-{{2t_1^2} \over U} \langle e^{ {i \over 2} \hat{\chi}_j}e^{ {i \over 2} \hat{\chi}_k}(c_{j \downarrow}c_{k \uparrow}
     -c_{j \uparrow}c_{k \downarrow} )\rangle= \Delta_{jk}
     \label{Delta}
  \end{eqnarray}
 exists.
The parameter $U$ is the on-site Coulomb repulsion parameter, and $t_1$ is the nearest neighbor hopping parameter in the CuO$_2$ plane. A salient feature of the above pair potential is that it is obtained using the particle-number fixed ground state. The surface bilayer Hamiltonian, $H^{\rm HF}_{\rm sbilayer}$, also includes the second nearest neighbor hopping with the hopping parameter $t_2$. 
The Hamiltonian $H^{\rm HF}_{\rm bbilayer}(\ell_b)$ is the $\ell_b$th bulk bilayer Hamiltonian.

Each CuO$_2$ plane extends in the $xy$ plane. The two layers in a bilayer are connected by the hopping in the $z$-direction. For example, $H^{\rm HF}_{\rm sbilayer}$ consists of two $H^{\rm HF}_{\rm surf}$ in Ref.~\cite{Koizumi2022aa} connected by the following interlayer Hamiltonian
  \begin{eqnarray}
   &&H^{\rm HF}_{\rm interlayer-ss}
   =
  -t_{ss}\sum_{\langle k_{s1},j_{s2} \rangle_{z}, \sigma}\left[\left(1-\langle n_{k_{s1}, -\sigma} \rangle \right)c^\dagger_{k_{s1}\sigma}c_{j_{s2}\sigma}\left(1-\langle n_{j_{s2}, -\sigma} \rangle \right)+{\rm H.c.} \right]
         \nonumber
       \\
  &+&{{2t_{ss}^2} \over U} \sum_{\langle k_{s1},j_{s2} \rangle_{z}}\Big[ \left( \langle { S}^x_{k_{s1}} \rangle -i   \langle { S}^y_{k_{s1}} \rangle \right) c^{\dagger}_{j_{s2} \uparrow} c_{j_{s2} \downarrow}+ \left(\langle { S}^x_{k_{s1}} \rangle + i   \langle { S}^y_{k_{s1}} \rangle \right) c^{\dagger}_{j_{s2} \downarrow} c_{j_{s2} \uparrow}
  \nonumber
  \\
  &+&
 \left( \langle { S}^z_{k_{s1}} \rangle-{1 \over 2} \langle n_{k_{s1}} \rangle \right)c^{\dagger}_{j_{s2} \uparrow} c_{j_{s2} \uparrow}- \left( \langle { S}^z_{k_{s1}} \rangle+{1 \over 2} \langle n_{k_{s1}} \rangle \right)c^{\dagger}_{j_{s2} \downarrow} c_{j_{s2} \downarrow} \Big]
  \nonumber
  \\
    &+&{{2t_{ss}^2} \over U} \sum_{\langle k_{s1},j_{s2} \rangle_{z}}\Big[ \left( \langle { S}^x_{j_{s2}} \rangle -i   \langle { S}^y_{j_{s2}} \rangle \right) c^{\dagger}_{k_{s1} \uparrow} c_{k_{s1} \downarrow}+ \left(\langle { S}^x_{j_{s2}} \rangle + i   \langle { S}^y_{j_{s2}} \rangle \right) c^{\dagger}_{k_{s1} \downarrow} c_{k_{s1} \uparrow}
  \nonumber
  \\
  &+&
 \left( \langle { S}^z_{j_{s2}} \rangle-{1 \over 2} \langle n_{j_{s2}} \rangle \right)c^{\dagger}_{k_{s1} \uparrow} c_{k_{s1} \uparrow}- \left( \langle { S}^z_{j_{s2}} \rangle+{1 \over 2} \langle n_{j_{s2}} \rangle )c^{\dagger}_{k_{s1} \downarrow} c_{k_{s1} \downarrow}\right) \Big]
  \nonumber
  \\
  &-&{{4t_{ss}^2} \over U} \sum_{\langle k_{s1},j_{s2} \rangle_{z}}\left( \langle {\bf S}_{k_{s1}}\rangle \cdot \langle {\bf S}_{j_{s2}}\rangle -{1 \over 4}\langle n_{i_{s1}}\rangle \langle n_{j_{s2}} \rangle \right)
  \label{H-interlayer}
  \end{eqnarray}
where $k_{s1}$ and $j_{s2}$ denote sites in the connected surface layers (denoted by $s1$ and $s2$), respectively; $\langle k_{s1},j_{s2} \rangle_{z}$ indicates a pair of sites connected by the $z$-direction hopping.  In a similar manner, $H^{\rm HF}_{\rm bilayer}(\ell_b)$ consists of two $H^{\rm HF}_{\rm bulk}(\ell_b$) (its explicit form is given by $H^{\rm HF}_{\rm EHFS}$ in Ref.~\cite{Koizumi2022aa}) connected by the above interlayer Hamiltonian with replacing surface sites $k_s, j_s$ by bulk sites $k_b, j_b$, and transfer integral $t_{ss}$ by $t_{bb}$.
Expectation values $\langle n_{k} \rangle$ and $\langle n_{k, \sigma} \rangle$ in Eq.~(\ref{H-interlayer}) are the mean values of the electron density and that with specific spin $\sigma$ ($-\sigma$ is the opposite component) at the $k$th site, respectively.
The expectation value $\langle S^\alpha_{k} \rangle$ indicates the mean value of the spin component $\alpha$ at the $k$th site,
and  $\langle {\bf S}_{k}\rangle \cdot \langle {\bf S}_{j}\rangle$ is given by $\langle S^x_{k}\rangle \cdot \langle S^x_{j}\rangle +\langle S^y_{k}\rangle \cdot \langle S^y_{j}\rangle +\langle S^z_{k}\rangle \cdot \langle S^z_{j}\rangle$.
The Hamiltonian $H^{\rm HF}_{\rm surface-bulk}$ is the interlayer Hamiltonian between the surface layer and the first bulk layer, and $H^{\rm HF}_{\rm bulk-bulk}(\ell_b, \ell_b +1)$ 
  is the interlayer Hamiltonian between the $\ell_b$th and $(\ell_b+1)$th bulk layers given in Ref.~\cite{Koizumi2022aa}. The parameters $h_{k \sigma, j \sigma'}$ are collections of contributions from all the terms that appear as parameters for the products of electron creation and annihilation operators $c_{k \sigma}^{\dagger} c_{j \sigma'}$.
  
Since the field operator $\hat{\Psi}_{\sigma}({\bf r})$ annihilates an electron with spin $\sigma$ at the spacial position ${\bf r}$, relations between the electron annihilation operators at site $k$ (its coordinate is ${\bf r}_k$), and the particle-number conserving Bogoliubov operators are deduced as
 \begin{eqnarray}
 c_{ k \uparrow} &=&\sum'_{n} [ u^{n}_{k \uparrow}\gamma_{n}- (v^{n}_{k \uparrow})^{\ast}\gamma_{n}^{\dagger}] e^{ -{i \over 2} \hat{\chi}_k}
 \nonumber
 \\
 c_{ k \downarrow} &=&\sum'_{n} [ u^{n}_{k \downarrow}\gamma_{n}+ (v^{n}_{k \downarrow})^{\ast}\gamma_{n}^{\dagger}] e^{ -{i \over 2} \hat{\chi}_k}
 \label{eq137}
  \end{eqnarray}
  from the field operators in Eq.~(\ref{eq1}),
  where $u^{n}_{k \sigma}$ and  $v^{n}_{k \sigma}$ correspond to $u^{n}_{\sigma}({\bf r}_k)$ and  $v^{n}_{\sigma}({\bf r}_k)$, respectively; here, $\sum'_{n}$ indicates  that the sum over $n$ only includes the positive Bogoliubov energy states.
  We choose the Bogoliubov operators to satisfy  the requirement in Eq.~(\ref{eq2}),
   and also 
$H_{\rm eff}$ is expressed as
    \begin{eqnarray}
H_{\rm eff}=\sum_{n}'  E_n  \gamma^{\dagger}_{n} \gamma_{n} + E'_{\rm const.}
\label{Heff-gamm}
  \end{eqnarray}
where $E'_{\rm const.}$ is a constant. Only the Bogoliubov energies with $E_n >0$ should be included since the Bogoliubov operators express the excitations from the ground state.

Using Eqs.~(\ref{Heff}), (\ref{eq137}), and (\ref{Heff-gamm}), the following PNC-BdG equations are obtained,
    \begin{eqnarray}
 E_{n} u^{n}_{k \uparrow}&=& \sum_{j \sigma'}  e^{ {i \over 2} \hat{\chi}_k} h_{k \uparrow, j \sigma'} e^{- {i \over 2} \hat{\chi}_j} u^{n}_{j \sigma'} + 
 \sum_j \Delta_{kj} v^{n}_{j \downarrow} 
  \nonumber
 \\
  E_{n} u^{n}_{k \downarrow}&=& \sum_{j \sigma'}  e^{ {i \over 2} \hat{\chi}_k} h_{k \downarrow, j \sigma'} e^{- {i \over 2} \hat{\chi}_j} u^{n}_{j \sigma'} + \sum_j \Delta_{ji} v^{n}_{j \uparrow} 
  \nonumber
 \\
  E_{n} v^{n}_{k \uparrow}&=& -\sum_{j } e^{ -{i \over 2} \hat{\chi}_k} h^{\ast}_{k \uparrow, j \uparrow} e^{ {i \over 2} \hat{\chi}_j} v^{n}_{j \uparrow} +\sum_{j}  e^{ -{i \over 2} \hat{\chi}_k} h^{\ast}_{k \uparrow, j \downarrow} e^{ {i \over 2} \hat{\chi}_j} v^{n}_{j \downarrow} + \sum_j \Delta_{kj}^{\ast} u^{n}_{j \downarrow} 
  \nonumber
 \\
  E_{n} v^{n}_{k \downarrow}&=& \sum_{j} e^{ -{i \over 2} \hat{\chi}_k} h^{\ast}_{k \downarrow, j \uparrow} e^{ {i \over 2} \hat{\chi}_j} v^{n}_{j \uparrow} -\sum_{j} e^{ -{i \over 2} \hat{\chi}_k} h^{\ast}_{k \downarrow, j \downarrow} e^{ {i \over 2} \hat{\chi}_j} v^{n}_{j \downarrow} +\sum_j \Delta_{jk}^{\ast} u^{n}_{j \uparrow} 
  \nonumber
  \\
  \end{eqnarray}
  By self-consistently solving the above equations, we can obtain $E_n, u^{n}_{j \sigma}$, and $v^{n}_{j \sigma}$. The self-consistent solution yields the $\chi_k=0$ case with zero current.
  The spin-texture is obtained from this solution. 
  
  \begin{figure}
\begin{center}
\includegraphics[width=9.0cm]{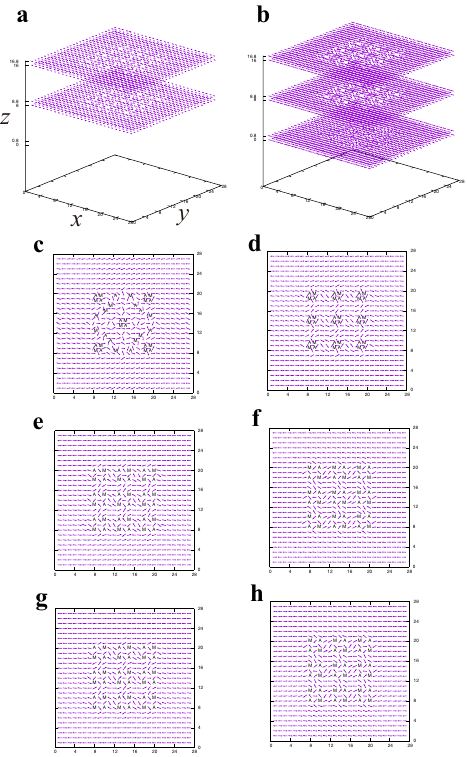}
\end{center}
\caption{Spin texture for a thin film composed of surface bilayer and two-bulk bilayers. Each bulk layer
contains nine $4 \times 4$ SVQs in a $27 \times 27$ square lattice. {\bf a}) Over all spin texture. {\bf b}) The normalized overall spin texture. The spin texture is the same as in ${\bf a}$ but normalized, separately, for each layer, for visualization purpose.
{\bf c}) Normalized spin texture for the lower ($z=0.0$) surface layer. ``M'' and ``A'' indicate centers of winding number $+1$ spin-vortex and winding number $-1$ spin-vortex, respectively. 
{\bf d}) Normalized spin texture for the upper ($z=0.8$) surface layer. 
{\bf e}) Normalized spin texture for the lower ($z=8.0$) layer of the first bulk bilayer.
{\bf f}) Normalized spin texture for the upper ($z=8.8$) layer of the first bulk bilayer.
{\bf g}) Normalized spin texture for the lower ($z=16.0$) layer of the second bulk bilayer.
{\bf h}) Normalized spin texture for the upper ($z=16.8$) layer of the second bulk bilayer.
}
\label{27spin}
\end{figure}

\begin{figure}
\begin{center}
\includegraphics[width=9.0cm]{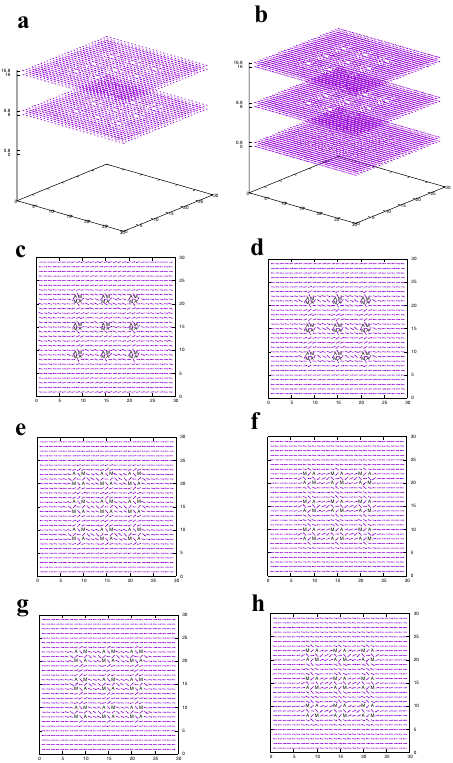}
\end{center}
\caption{The same as Fig.~\ref{27spin} but each bulk layer
contains nine  $4 \times 4$ SVQs in a $29 \times 29$ square lattice.}
\label{29spin}
\end{figure}
  
 Let us examine the spin texture obtained. Results are depicted in Figs.~\ref{27spin} and \ref{29spin}.
 The parameters for the model Hamiltonian are following: the first nearest neighbor hopping parameter in the CuO$_2$ plane, $t_1$, is taken to be $t_1=0.2$ eV. The second nearest neighbor hopping parameter in the CuO$_2$ plane, $t_2$, is  $t_2 =-0.12t_1$. The on-site Coulomb repulsion parameter, $U$, is $U = 8t_1$. The Rashba spin-orbit interaction parameter, $\lambda$, is $\lambda= 0.02t_1$. The hopping parameter between the surface and bulk layers, $t_{sb}$, is $t_{sb} = 0.01t_1$. The hopping parameters between layers in a bilayer, $t_{ss}$ for the surface bilayer, $t_{bb}$ for the bulk bilayer, are $t_{ss}=t_{bb} = 0.2t_1$. The bulk chemical potential, $\mu_{bulk}$, and surface chemical potential, $\mu_{surf}$,  are $\mu_{bulk} = 4t_1$ and $\mu_{surf} =-0.1t_1$, respectively. Please consult Ref.~\cite{Koizumi2022aa} for the details of those parameters.
 
 The unit of length is the CuO$_2$ plane lattice constant $a \approx 0.4$ nm. The surface of the Bi-2212 thin film is located at $z=0.0$, and the thin film exists in $z \ge 0$. The surface CuO$_2$ layers of the bilayer exist at $z=0.0$ and $z=0.8$; the first bulk CuO$_2$ layers of the bilayer exist at $z=8.0$ and $z=8.8$; and the second bulk CuO$_2$ layers exist at $z=16.0$ and $z=16.8$.

The calculation is done with the open boundary condition with a peripheral antiferromagnetic region; this region is necessary to obtain numerically converged results; without this region, the self-consistent calculations do not converge.
The spin-vortices are created by itinerant electrons with spin moments lying in the $xy$ plane. The expectation values of the components of the spin are given by
\begin{eqnarray}
 \langle { S}^x_k\rangle +i   \langle { S}^y_{k} \rangle=S_k e^{i \xi_k}, \  \langle S^z_k \rangle=0
\end{eqnarray}
where $S_k$ is a real number, and  $\xi_k$ is the polar angle of the spin at the $k$th site.
Each spin-vortex is characterized by the winding number defined by
\begin{eqnarray}
w_{C_{\ell}}[\xi] ={ 1 \over {2 \pi}}\oint_{C_\ell} \nabla \xi \cdot d {\bf r}
\label{w-xi}
\end{eqnarray}
It is calculated for a loop formed by surrounding lattice sites around each center of small polaron.
If the winding number is $+1$, it is called a `meron' and  denoted by ``M'', and  if it is $-1$,
it is called an `antimeron' and denoted by ``A'' in  Fig.~\ref{27spin}. 

The winding numbers for the bulk spin-vortices are input parameters for the self-consistent calculation. The combination of four-vortex unit is found energetically favorable in our previous works, and called the `spin-vortex-quartet (SVQ)'. The input spin winding numbers are so arranged that SVQs are placed periodically.
On the other hand, the spin vortices in the surface layers are obtained without winding number inputs;
they are automatically formed through the interaction from the bulk layers.

In Fig.~\ref{27spin}, the result for the lattice with $27 \times 27$ CuO$_2$ planes is depicted.
Spin vortices are formed around small polarons in the bulk layers.
The spin-texture in the bulk layers are those arising from $4 \times 4$ SVQs with their winding numbers 
equal to the input winding numbers.
Each bulk layer is in the state of the `effectively-half-filled-situation', where the number of electrons and that for the sites allowed for electron hopping are the same.
The bulk chemical potential, $\mu_{bulk}$, is chosen to have this effectively-half-filled-situation.
The surface chemical potential, $\mu_{surf}$, is so chosen that the number of electrons in the surface layer is almost equal to that of the bulk layer.
Spin-vortices are also generated in the surface bilayers although small polarons are absent.
The spin-texture in the lower ($z=0.0$) layer shows rather frustrated spin arrangement compared with the upper ($z=0.8$) layer. 

In Fig.~\ref{29spin}, the result is depicted
for the lattice with $29 \times 29$ CuO$_2$ planes.
The frustrated spin-texture existed in the lower ($z=0$) layer of the surface layer in 
 Fig.~\ref{27spin} is absent. We consider SVILCs arising from the above two different spin arrangements in the following.

\section{Spin-vortex-induced loop currents (SVILCs) and generated magnetic fields by them}
\label{sec4}
  
The self-consistent solutions $u^{n}_{j \sigma}$, and $v^{n}_{j \sigma}$ may not be valid ones when spin-vortices exist since they may be multi-valued functions of the coordinate $j$.
Let us see this point below:
for example, the term like $\left(\langle { S}^x_{k_{s1}} \rangle -i   \langle { S}^y_{k_{s1}} \rangle \right) c^{\dagger}_{j_{s2} \uparrow} c_{j_{s2} \downarrow}$ in Eq.~(\ref{H-interlayer}) shows $\xi$ dependence as
\begin{eqnarray}
e^{-i\xi_{k_{s1}}}c^{\dagger}_{j_{s2} \uparrow} c_{j_{s2} \downarrow}
\end{eqnarray}
and $\left(\langle { S}^x_{k_{s1}} \rangle + i   \langle { S}^y_{k_{s1}} \rangle \right) c^{\dagger}_{j_{s2} \downarrow} c_{j_{s2} \uparrow}$
as 
\begin{eqnarray}
e^{i\xi_{k_{s1}}}c^{\dagger}_{j_{s2} \downarrow} c_{j_{s2} \uparrow}
\end{eqnarray}
Considering that the expectation values of the above terms should depend on the difference of the phases $\xi_{k_{s1}}-\xi_{j_{s2}}$, we have
\begin{eqnarray}
\langle c^{\dagger}_{j_{s2} \uparrow} c_{j_{s2} \downarrow} \rangle \sim e^{i\xi_{j_{s2}}},
\quad 
\langle c^{\dagger}_{j_{s2} \downarrow} c_{j_{s2} \uparrow} \rangle \sim e^{-i\xi_{j_{s2}}},
\end{eqnarray}
Taking into account  the relations in Eq.~(\ref{eq137}), $\xi_j$ dependencies
in $u_{j \sigma}^n$ and $v_{j \sigma}^n$ are deduced as
\begin{eqnarray}
u_{j \uparrow}^n, v_{j \downarrow}^n \sim  e^{-{i \over 2}\xi_j}, \quad u_{j \downarrow}^n, v_{j \uparrow}^n \sim  e^{{i \over 2}\xi_j}, 
\end{eqnarray}
When an excursion of the value of $\xi$ is performed starting from $\xi_j$ around a loop in the coordinate space, $\xi_j$ may become $\xi_j + 2\pi k$ after the one around, where $k$ is an integer;
 if $k$ is odd,
 $e^{\pm{i \over 2}\xi_j}$ becomes  $-e^{\pm{i \over 2}\xi_j}$ after the excursion,
 yielding the multi-valuedness.
In order to achieve the single-valued constraint with respect to the coordinate, we use $\chi_j$ dependency. Including it, the overall phase factors become
\begin{eqnarray}
u_{j \uparrow}^n, u_{j \downarrow}^n \sim  e^{-{i \over 2}\chi_j} e^{-{i \over 2}\xi_j}, \quad u_{j \downarrow}^n, u_{j \uparrow}
^n \sim  e^{-{i \over 2}\chi_j}e^{{i \over 2}\xi_j}
\end{eqnarray}
and the multi-valuedness is avoided if $\chi_i$ is chosen so that the following requirement is fulfilled,
 \begin{eqnarray}
w_{C_{\ell}}[\xi]+ w_{C_{\ell}}[\chi] = \mbox{even number}  \mbox{  for any loop $C_{\ell}$}
\label{eqSingle}
\end{eqnarray}
where $w_{C_{\ell}}[\chi]$ is the winding number for $\chi$ given by
\begin{eqnarray}
w_{C_{\ell}}[\chi] = { 1 \over {2 \pi}}\oint_{C_\ell} \nabla \chi \cdot d {\bf r}
\end{eqnarray}
We obtain $\chi_j$ by requiring the above conditions for independent loops $\{ C_1, \cdots, C_{N_{\rm loop}} \}$ formed as the boundaries of plaques of the flattened lattice.
Here, 
$N_{\rm loop}$ is the total number of plaques of the flattened lattice, and the flattened lattice is a 2D lattice constructed from the original 3D lattice by removing some faces and walls. Consult Ref.~\cite{koizumi2022b} for detail.

 Actually, what we need to know is the differences of the phases, $(\chi_i - \chi_j)$'s, between bonds connecting sites. Therefore, the number of unknowns to be evaluated is equal to the number of the bonds, $N_{\rm bond}$.
The winding number requirement provides $N_{\rm loop}$ equations. However,
$N_{\rm loop}$ equations are not enough to obtain all $(\chi_i - \chi_j)$'s.
We add the conservation of the local charge requirements, which gives rise to the number of sites (we denote it as $N_{\rm site}$) minus one conditions, where minus one comes from the fact that the total charge is conserved in the calculation with a fixed number of electrons.
Then, the solvability condition for all $(\chi_i - \chi_j)$'s is
\begin{eqnarray}
N_{\rm bond}=N_{\rm loop}+N_{\rm site}-1
\label{Euler1}
\end{eqnarray}
This condition is satisfied since it agrees with the Euler's theorem for 2D lattice given by
\begin{eqnarray}
[\mbox{\# edges}]=[\mbox{\# faces}]+[\mbox{\# vertices}-1]
\label{Euler2}
\end{eqnarray}
where `$\mbox{\# A}$' means `the number of $A$'.

\begin{figure}[H]
\begin{center}
\includegraphics[width=9.0cm]{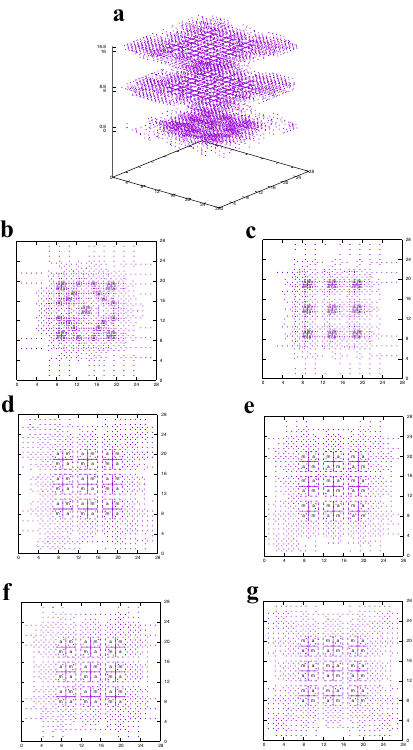}
\end{center}
\caption{Current distributions for the system with spin distribution shown in Fig.~\ref{27spin}. The  winding numbers of the SVILCs are assumed to be the same as the underlying SVQs winding numbers.
{\bf a}) Overall current distributions for the system. 
{\bf b}) Current distribution for the lower ($z=0.0$) surface layer. ``m'' and ``a'' indicate centers of winding number $+1$ SVILC and winding number $-1$ SVILC, respectively. 
{\bf c}) Current distribution for the upper ($z=0.8$) surface layer. 
{\bf d}) Current distribution for the lower ($z=8.0$) of the first bulk bilayer.
{\bf e}) Current distribution for the upper ($z=8.8$) layer of the first bulk bilayer.
{\bf f}) Current distribution for the lower ($z=16.0$) layer of the second bulk bilayer.
{\bf g}) Current distribution for the upper ($z=16.8$) layer of the second bulk bilayer.
}
\label{27_current_maam-2}
\end{figure}

In Fig.~\ref{27_current_maam-2}, a current distribution generated by the SVILCs for the spin texture given in Fig.~\ref{27spin} is depicted.
In this current distribution, the  winding numbers of the SVILCs are assumed to be the same as the underlying SVQs winding numbers; from our previous calculation experience in similar systems, we noticed that this case will be the lowest energy state. 
Although the magnitude of the moments is very small, spin-vortices are also formed in the surface bilayer by the influence of the bulk bilayers as seen in  Fig.~\ref{27spin}.
Therefore, they also induce SVILCs with the magnitude comparable to those generated in the bulk layers as seen in Figs.~\ref{27_current_maam-2} {\bf b} and {\bf c}. 
However, the magnetic field generated is too small to be used as the detection of the SVILCs in this case.
 
\begin{figure}[H]
\begin{center}
\includegraphics[width=12.0cm]{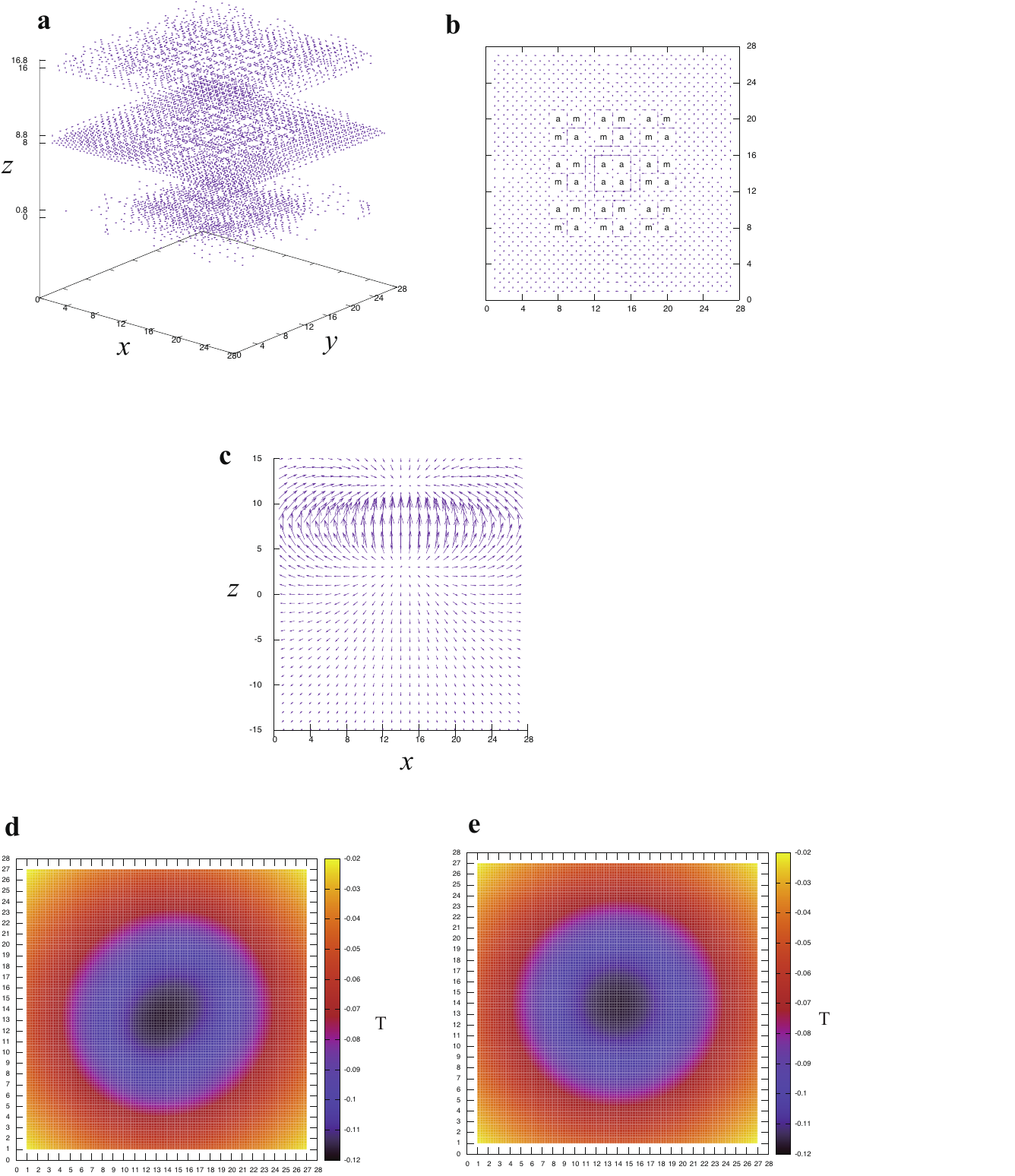}
\end{center}
\caption{Current distributions for the system with spin distribution shown in Fig.~\ref{27spin} with modified current distribution in one layer from {\bf d} in Fig.~\ref{27_current_maam-2} to {\bf b} in this figure.
{\bf a}) Overall current distribution for the system. 
{\bf b}) Modified current distribution for the lower ($z=8.0$) layer of the first bulk bilayer.
{\bf c}) Magnetic field generated by SVILCs in the $xz$ plane.
{\bf d}) The $z$ component of the magnetic field at $z=-10.0$. 
{\bf e}) The $z$ component of the magnetic field at $z=-10.0$ calculated with subtracting the surface current contribution.
}
\label{27-aaaa-2}
\end{figure}

The generated magnetic field by SVILCs can be increased by changing winding numbers of the SVILCs.
In Fig.~\ref{27-aaaa-2}, the SVILC in the $z=8.0$ layer is modified from {\bf d} in Fig.~\ref{27_current_maam-2} to {\bf b} in Fig.~\ref{27-aaaa-2}. 
The magnetic field generated in the $xz$ plane at $y=0$ is depicted Fig.~\ref{27-aaaa-2}{\bf c}.
The contour plot of the $z$-component of the magnetic field at $z=-10.0$ is shown in Fig.~\ref{27-aaaa-2}{\bf d}. The magnitude of the magnetic field is around $0.1$T, thus, it will be large enough to be detected. 
The SVILC combination of Fig.~\ref{27-aaaa-2}{\bf b} may be realized by exciting the loop current state by applying a pulsed magnetic field. If the aimed loop pattern is excited by chance, and the resulting state stays for a time long enough to be detected, we can confirm the existence of the SVILCs from 
the generated magnetic field detection.
The surface current contribution to the magnetic field is very small compared to the contribution from
the current in Fig.~\ref{27-aaaa-2}{\bf b}. In Fig.~\ref{27-aaaa-2}{\bf e},
the magnetic field with subtracting the surface current contribution is depicted. It is almost the same as the one
in Fig.~\ref{27-aaaa-2}{\bf d}, indicating that the surface current contribution is negligible.

\begin{figure}[H]
\begin{center}
\includegraphics[width=9.0cm]{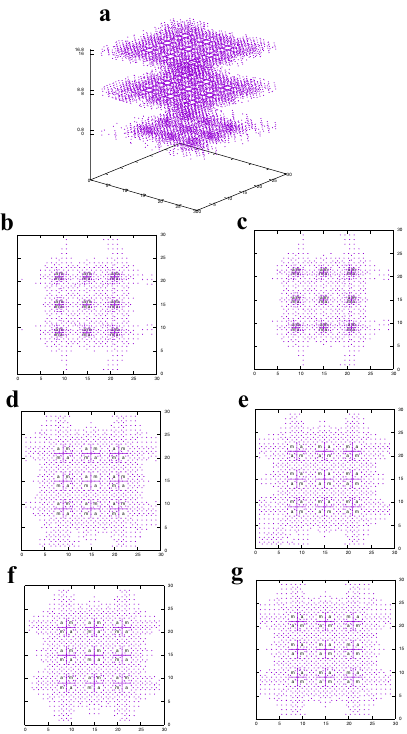}
\end{center}
\caption{The same as Fig.~\ref{27_current_maam-2} but with spin distribution shown in Fig.~\ref{29spin}.}
\label{29_current_maam-2}
\end{figure}

In Fig.~\ref{29_current_maam-2}, a current distribution generated by the SVILCs for the spin texture given in Fig.~\ref{29spin} is depicted.
The winding numbers of the SVILCs  are the same as the underlying SVQs winding numbers. 
The magnetic field generated is too small to be used as the confirmation for the existence of the SVILCs  in the case, too.

\begin{figure}[H]
\begin{center}
\includegraphics[width=12.0cm]{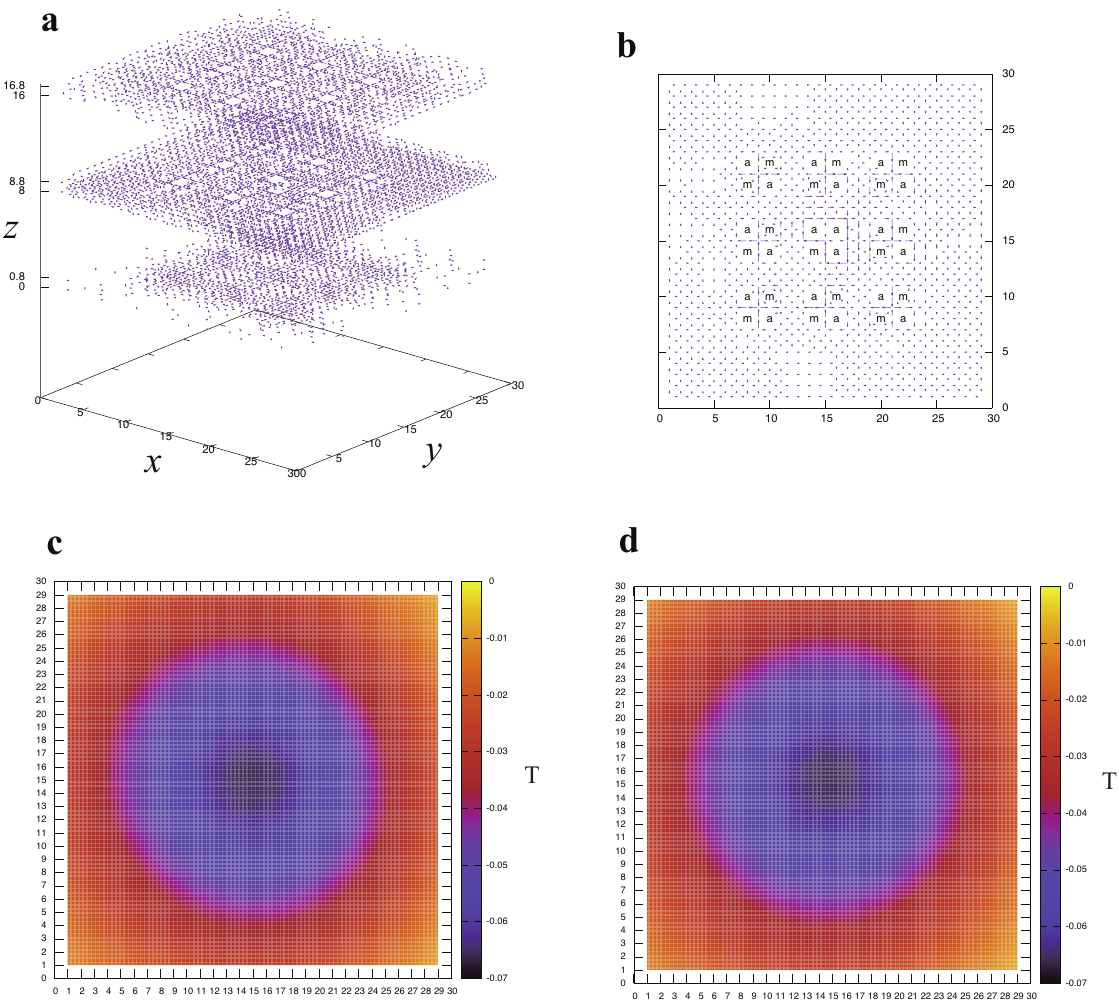}
\end{center}
\caption{Current distributions for the system with spin distribution shown in Fig.~\ref{29spin} with 
modified current distribution in one layer from {\bf d} in Fig.~\ref{29_current_maam-2} to {\bf b} in this figure.
{\bf a}) Overall current distribution for the system. 
{\bf b}) Modified current distribution for the lower ($z=8.0$) layer of the first bulk bilayer.
{\bf c}) The $z$ component of the magnetic field at $z=-10.0$. 
{\bf d}) The $z$ component of the magnetic field at $z=-10.0$ calculated with subtracting the surface current contribution.
}
\label{29-maaa-2}
\end{figure}

In order to increase the magnetic field, we change the current pattern of a SVQ.
The change made for the previous case (see Fig.~\ref{27-aaaa-2}{\bf b}), actually, cannot be obtained in this system since the numerical calculation does converge.
Instead, the SVILC in the layer at $z=8.0$ is modified to Fig.~\ref{29-maaa-2}{\bf b}. 
In this case, a large magnetic field that may be detectable is obtained. The magnitude is smaller than the
one obtained in Fig.~\ref{27-aaaa-2}, but still it is in the order of mT, thus, will be detectable.
This may be achieved by exciting the loop current state by applying a pulsed magnetic field.
Note that this current pattern change does not give a converged result for the spin texture in Fig.~\ref{27spin}. Our calculations indicate that the stable current patters depend on the underlying spin texture.

\begin{figure}[H]
\begin{center}
\includegraphics[width=12.0cm]{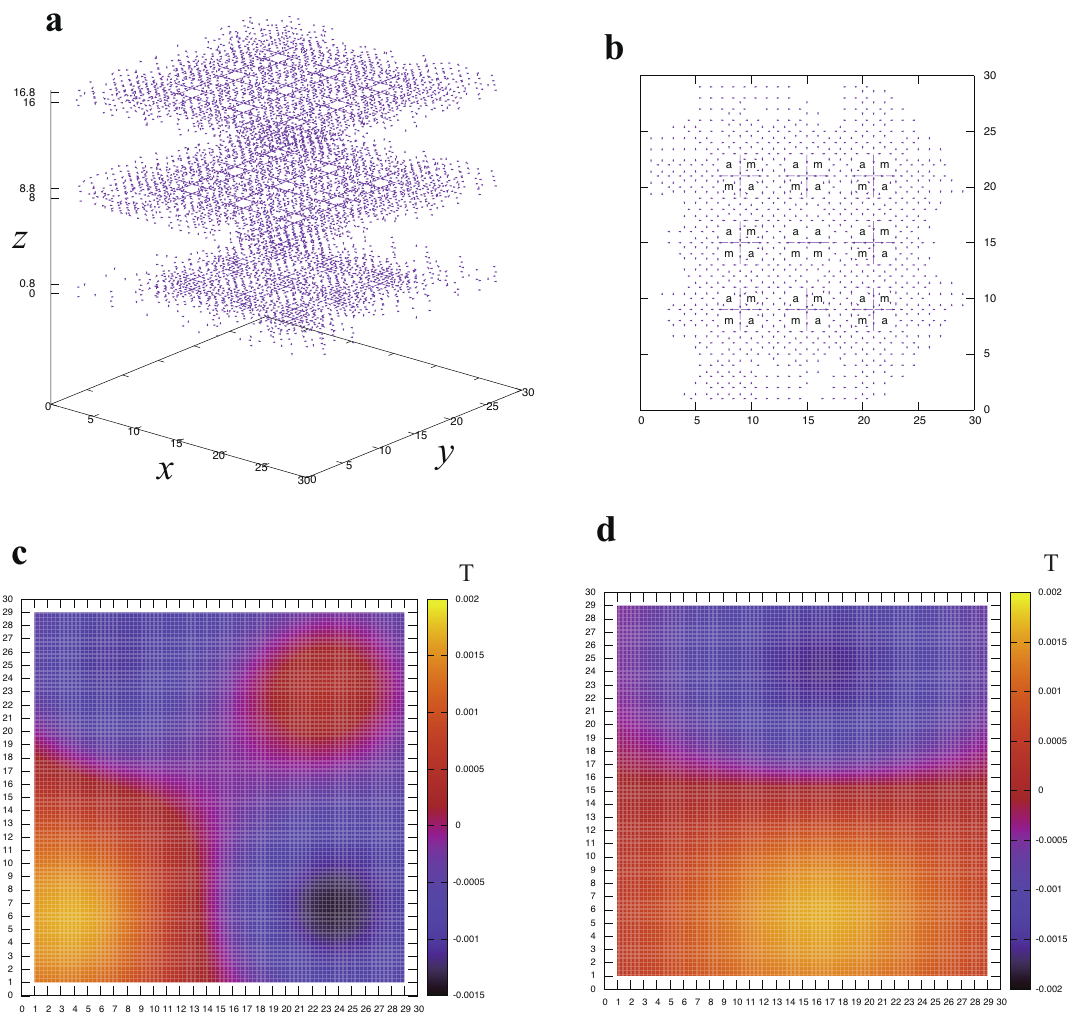}
\end{center}
\caption{Current distributions for the system with spin distribution shown in Fig.~\ref{29spin} with 
modified current distribution in one layer from {\bf d} in Fig.~\ref{29_current_maam-2} to {\bf b} in this figure.
{\bf a}) Overall current distribution for the system. 
{\bf b}) Modified current distribution for the lower ($z=8.0$) layer of the first bulk bilayer.
{\bf c}) The $z$ component of the magnetic field at $z=-10.0$. 
{\bf d}) The $z$ component of the magnetic field at $z=-10.0$ calculated with subtracting the surface current contribution.
}
\label{29-mama-2}
\end{figure}

We also calculate a different current pattern depicted in Fig.~\ref{29-mama-2} by
modifying the SVILCs in the layer at $z=8$ to Fig.~\ref{29-mama-2}{\bf b}. 
In this case the surface current contribution is comparable to the magnetic field generated by 
the SVULCs in the bulk layer. If the surface contribution is subtracted as shown in Fig.~\ref{29-mama-2}{\bf d}, 
a clear contribution from the modified SVILCs in the bulk is obtained. 
This current state is the one we assumed as a qubit state \cite{Koizumi:2022aa},
which is energetically closer to the one in Fig.~\ref{29_current_maam-2}, compared to other SVILC excited states considered in this work.
Note that this current pattern change does not give a converged result for the spin texture in Fig.~\ref{27spin}.

\section{Concluding remarks}
\label{sec5}
 
The existence of nano-sized loop currents in the cuprate is supported by many experiments.
The existence of spin-vortices is confirmed by experiments although their size is much larger than the one considered in this work \cite{Wang:2023aa}; however, if the experimental resolution is improved, the predicted nano-sized spin-vortices considered in this work may also be observed.

It is important to check if the nano-sized loop currents in the cuprate are whether the SVILCs or not.
The present work indicates that if they are the SVILCs they will produce detectable magnetic fields
depending on the current patterns. Since the spin moments of spin-vortices are lying in the CuO$_2$ planes, their magnetic field does not have the component perpendicular to the CuO$_2$ planes.
On the other hand, the magnetic field produced by the SVILCs has the perpendicular component,
thus, both the spin-vortices and SVILCs can be separately detected.
If their existence is confirmed, it will lead to the elucidation of the mechanism of the cuprate superconductivity. Further, it may provide new qubits that can realize practical quantum computers. 

\bibliographystyle{spphys}% APS-like style for physics

% \bibliography{SPIN-BCS}
 
\end{document}